\documentclass[aps,prl,twocolumn,amsmath,amssymb,superscriptaddress,nofootinbib,10pt]{revtex4-2}

\usepackage{bm}
\usepackage{physics}
\usepackage{graphicx}
\usepackage{microtype}
\usepackage{hyperref}
\allowdisplaybreaks[1]

\begin{document}

\title{Divergent Fluctuations from a 2D Infrared Catastrophe}

\author{Richard G. Hennig}
\affiliation{Department of Materials Science, University of Florida, Gainesville, FL 32611, USA}
\affiliation{Quantum Theory Project, University of Florida, Gainesville, FL 32611, USA}

\author{Clotilde S. Cucinotta}
\affiliation{Department of Chemistry, Imperial College London, London, W12 0BZ, United Kingdom}

\date{\today}

\begin{abstract}
Molecular simulations of interfacial polar media routinely employ periodic boundary conditions parallel to the interface. We show that this lateral periodicity introduces a spatially uniform in-plane mode ($q_{\parallel}=0$) that is unscreened because every lateral replica carries identical charge fluctuations. This 2D mode reduces the plane-averaged potential to a stochastic integral of the plane-averaged charge density along $z$, so that in a semi-infinite slab the variance of the potential grows linearly with depth. In a finite or periodic cell along $z$, with boundaries held at fixed potential, it follows a parabolic profile--a Brownian bridge--pinned to zero at both ends, with amplitude inversely proportional to the lateral cell area. These diverging fluctuations are a pure artifact of the imposed 2D lateral periodicity: they remain bounded in systems that are non-periodic or of finite lateral extent. We provide an analytic expression for their magnitude in dipolar media, yielding a practical criterion for the choice of lateral cell dimensions.
\end{abstract}

\maketitle

Fluctuations of the electrostatic potential drive rates, selectivity, and stability across problems that span electrochemistry and molecular biology. In electrode–electrolyte systems, the variance and spectrum of the plane-averaged potential feed directly into key observables, including double-layer capacitance, the potential of zero charge, reorganization energies for electron transfer, ionic screening, interfacial and solvation free energies, and noise floors in nanoscale sensing~\cite{Kornyshev2007JPCB,FedorovKornyshev2014ChemRev,Merlet2012NatMater,Limmer2013PRL,BazantStoreyKornyshev2011PRL,Marcus1993RMP,Blumberger2015ChemRev,Fragasso2020ACSNano,Dekker2007NatNano}. In practice, these quantities are often obtained by laterally averaging charge or polarization and integrating the 1D Poisson equation in molecular simulations~\cite{GurtovenkoVattulainen2007, Spohr1997}.

In biomembranes and nanopores, the same plane-average workflow is used to reconstruct membrane potentials and gating charges~\cite{GurtovenkoVattulainen2007,Dekker2007NatNano,Fragasso2020ACSNano}. Similar analyses appear in ferroelectric and piezoelectric thin films, polar 2D materials, and molecular electronics, where macroscopic boundary conditions set device-relevant internal fields~\cite{StengelVanderbiltSpaldin2009NatPhys}. At electrified metal–electrolyte interfaces, recent \emph{ab initio} studies have resolved Pt–water double layers under bias~\cite{Khatib2021ElectrochimActa, DarbyCucinotta2022COELEC, Raffone2025CommChem} and developed constant-potential, open-boundary schemes for controlling electrode potential~\cite{Buraschi2024JPCL,Ahart2024JCTC}. In many of these settings, recent studies report large electrostatic potential fluctuations that persist over nanometer scales and argue that these fluctuations arise from collective electrostatic modes and from slow interfacial dynamics that directly influence the thermodynamics and kinetics of water and electrolytes~\cite{Deissenbeck2021PRL,Todorova2025Review}.

In this Letter we identify a catastrophe of the uniform plane mode ($q_{\parallel}=0$), inherent to the planar electrostatic Green's function of 2D-periodic slabs, that causes an \emph{apparent divergence} of plane-averaged potential fluctuations with slab thickness. Previous work on electrostatic boundary conditions and Poisson solvers in slab and interfacial geometries has focused primarily on obtaining accurate \emph{mean} electrostatic potentials, fields, and energies~\cite{YehBerkowitz1999,ToukmajiBoard1996,deLeeuwI1980,deLeeuwII1980,Chen2011PRB,OtaniSugino2006PRB}, whereas the impact of these choices on the \emph{moments and cumulants} of the potential has received far less systematic attention—even though the variance directly enters thermodynamic and kinetic quantities such as free energies and rates~\cite{Marcus1993RMP,Blumberger2015ChemRev}.

\begin{figure*}[t]
  \centering
  \includegraphics[width=\textwidth]{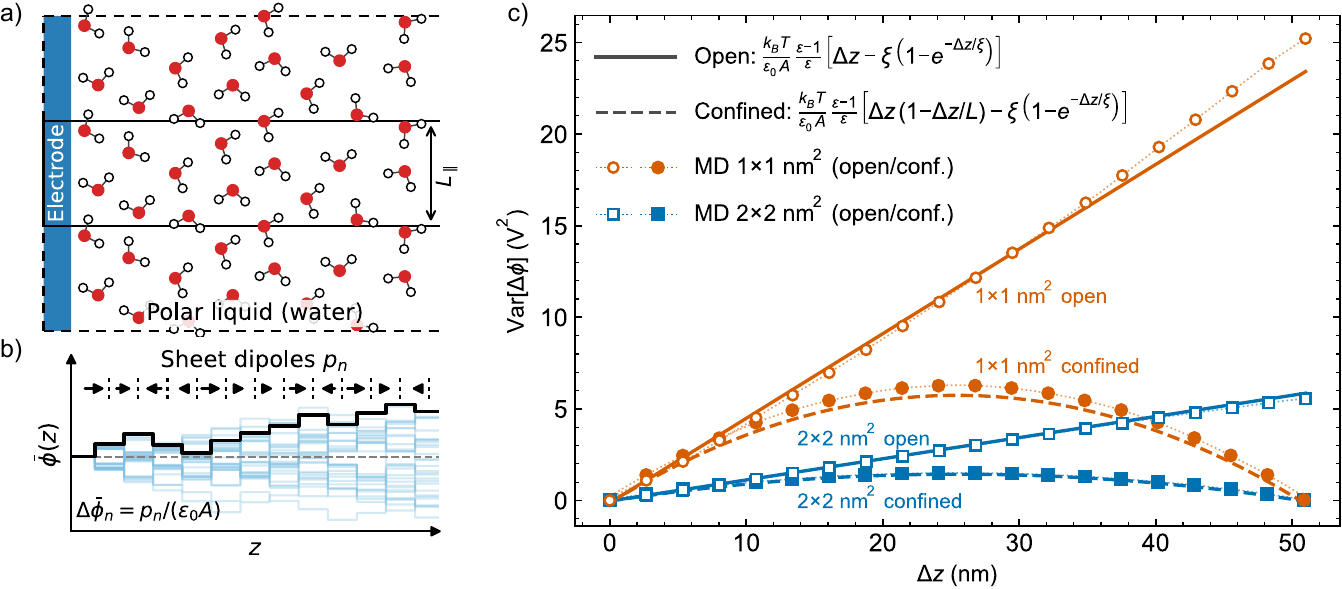}
  \caption{\label{fig:main}
    \textbf{Uniform plane mode induced variance growth under 2D periodicity.}
    (a)~2D-periodic slab geometry: an electrode (left) in contact with a polar liquid; the simulation cell is replicated in the interfacial plane (one in-plane dimension sketched with length $L_{\parallel}$ and area $A$).
    (b)~Coarse-grained $q_{\parallel}=0$ description along the surface-normal coordinate $z$: discretizing the plane-averaged polarization into sheet dipoles $p_n$ produces potential jumps $\Delta\bar{\phi}_n=p_n/(\varepsilon_0 A)$ and hence a cumulative step profile $\bar{\phi}(z)$ (thick line); light traces indicate representative stochastic realizations.
    (c)~Variance of the plane-averaged potential difference $\mathrm{Var}[\Delta\bar\phi(\Delta z)]$ for lateral cross sections $A=1\times1$ and $2\times2$~nm$^2$. Symbols show MD results for flexible TIP3P water in both single-interface (open, linear growth) and confined (Brownian-bridge) geometries. Solid curves are the parameter-free analytic model $V(\Delta z)=S[\Delta z - \xi(1-e^{-\Delta z/\xi})]$ for the open case and $V(\Delta z)=S[\Delta z(1-\Delta z/L) - \xi(1-e^{-\Delta z/\xi})]$ for the confined case, using $S=k_BT(\varepsilon{-}1)/(\varepsilon_0 A\varepsilon)$ with $\varepsilon=82$ and $\xi=0.3$~nm. The $1/A$ scaling is evident in both geometries; the analytic prediction matches the MD data with no fitted parameters.}
\end{figure*}

Because 2D periodicity forces every lateral replica to carry the same charge fluctuation, the uniform plane mode is entirely unscreened. It turns the plane-averaged electrostatic potential into a sum of weakly correlated potential jumps—a Wiener process in $z$—whose variance grows linearly with depth or parabolically across a finite slab, with a prefactor $\propto 1/A$. As with classic divergences that reflect mathematical idealization rather than physical reality—the ultraviolet catastrophe, the infrared divergence in QED, and Stokes' paradox—this growth is an artifact of boundary conditions, not emergent physics~\cite{Planck1901,Stokes1851}. We combine the analytic theory with MD simulations of water to validate these predictions and estimate the lateral cell areas needed to control the artifact. Figure~\ref{fig:main} illustrates the geometry and resulting variance divergence.

{\it Uniform plane mode and sheet-dipole mapping}---To obtain the scaling of plane-averaged potential fluctuations, we resolve Poisson's equation into lateral Fourier modes and isolate the \emph{uniform plane mode} ($\mathbf{q}_\parallel=0$). As shown below, this mode converts the plane-averaged potential into a cumulative integral of the polarization along $z$, producing a variance that grows with slab thickness. We write positions as $\mathbf{r}=(\mathbf{R},z)$ with lateral coordinate $\mathbf{R}\in\mathbb{R}^2$ and surface-normal coordinate $z$. The electrostatic potential $\phi$ obeys Poisson's equation,
\begin{equation}
-\varepsilon_0\,\nabla^2 \phi(\mathbf{R},z) = \rho(\mathbf{R},z),
\label{eq:Poisson3D}
\end{equation}
where $\varepsilon_0$ is the vacuum permittivity and $\rho$ the charge density.
We resolve the potential and charge by lateral two-dimensional Fourier transforms in $\mathbf{q}$ (hereafter $\mathbf{q}\equiv\mathbf{q}_\parallel$), without assuming any particular lateral boundary condition:
\begin{equation}
\begin{aligned}
\phi(\mathbf{R},z)
&= \int \!\frac{d^2\mathbf{q}}{(2\pi)^2}\,
   \phi_{\mathbf{q}}(z)\,e^{i\mathbf{q}\cdot\mathbf{R}},\\
\rho(\mathbf{R},z)
&= \int \!\frac{d^2\mathbf{q}}{(2\pi)^2}\,
   \rho_{\mathbf{q}}(z)\,e^{i\mathbf{q}\cdot\mathbf{R}}.
\end{aligned}
\label{eq:FourierExp}
\end{equation}
In a periodic cell the same formulas hold with $\mathbf{q}$ discretized. Inserting Eq.~\eqref{eq:FourierExp} into Eq.~\eqref{eq:Poisson3D} yields,
\begin{equation}
\phi_{\mathbf{q}}''(z) - |\mathbf{q}|^2\,\phi_{\mathbf{q}}(z) \;=\; -\,\rho_{\mathbf{q}}(z)/\varepsilon_0.
\label{eq:ODEmodes}
\end{equation}

The uniform plane mode ($\mathbf{q}=0$) yields the plane averages, $\bar\phi(z)\equiv \phi_{\mathbf{0}}(z)$ and $\bar \rho(z)\equiv \rho_{\mathbf{0}}(z)$, which obey
\begin{equation}
\bar\phi''(z) \;=\; -\bar \rho(z)/\varepsilon_0.
\label{eq:Poisson1D}
\end{equation}
Integrating this equation using the Green's function $G_0(z)=|z|/2$ for $d^2/dz^2$ on $\mathbb{R}$, we obtain
\begin{equation}
\bar\phi(z) = -\,\frac{1}{2\varepsilon_0}\!\int_{-\infty}^{\infty}\!|z-z'|\,\bar \rho(z')\,dz'
\;+\; E_0\,z + \phi_0,
\label{eq:phi_general_rho}
\end{equation}
with constants $E_0,\phi_0$ set by the macroscopic boundary conditions.

For any ${\bf q}\ne0$, the solution of Eq.~\eqref{eq:ODEmodes},
\begin{equation}
\phi_{\mathbf{q}}(z)
= \frac{1}{2\varepsilon_0|\mathbf{q}|}\int_{-\infty}^{\infty}
  e^{-|\mathbf{q}|\,|z-z'|}\,\rho_{\mathbf{q}}(z')\,dz' ,
\label{eq:phi_Gneq0}
\end{equation}
decays in $z$ with an associated decay length of $1/|\mathbf{q}|$. In a finite lateral cell the smallest nonzero lateral wavenumber scales as $|\mathbf{q}|_{\min}\sim 1/L_{\parallel}$, so the slowest nonuniform mode decays over a $z$‑scale on the order of $L_{\parallel}$ (e.g., $L_{\parallel}/2\pi$ under 2D periodicity). Since these nonuniform modes vanish when the potential is averaged over the full lateral periodic cell, only the $\mathbf{q}=0$ mode contributes to the plane-averaged potential. In what follows, we focus on systems simulated with 2D-periodic boundary conditions and analyze this mode; later, we compare with nonperiodic control geometries.

We decompose the total charge density into free charge $\rho_f$ and bound charge from the polarization density $\mathbf{P}$, $\rho(\mathbf{r}) = \rho_f(\mathbf{r}) - \nabla\!\cdot\!\mathbf{P}(\mathbf{r})$. Under full plane averaging over the periodic lateral cell, the lateral divergence terms $\partial_x P_x$ and $\partial_y P_y$ integrate to zero, so
\begin{equation}
    \overline{\nabla\cdot \mathbf{P}}(z)=\partial_z \bar P_z(z).
\end{equation}
Hence,
\begin{equation}
    \bar\rho(z)=\bar\rho_f(z)-\partial_z \bar P_z(z).
\end{equation}
In bulk water away from electrodes the plane-averaged free charge is negligible; we therefore set $\bar \rho_f(z)=0$ (a systematic treatment for free charges will be presented elsewhere; the general case is given in the Supplemental Material).

Substituting $\bar\rho(z')=-\partial_{z'}\bar P(z')$ into Eq.~\eqref{eq:phi_general_rho} and integrating by parts with respect to $z'$, one finds that the plane-averaged electric field satisfies
\begin{equation}
    \bar E(z)=-\partial_z \bar\phi(z)=-\frac{\bar P(z)}{\varepsilon_0},
\label{eq:Efield_main}
\end{equation}
up to constants fixed by the boundary conditions. Integrating once with respect to $z$ then gives
\begin{equation}
    \bar\phi(z)=\frac{1}{\varepsilon_0}\int^z \bar P(z')\,dz' + \mathrm{const}.
\label{eq:phi_cumulative}
\end{equation}
Thus the plane-averaged potential at depth $z$ is the cumulative integral of the plane-averaged polarization. A local potential difference over a distance $\Delta z$ is therefore
\begin{equation}
\Delta\bar\phi(z_0,\Delta z)\equiv \bar\phi(z_0+\Delta z)-\bar\phi(z_0)
=\frac{1}{\varepsilon_0}\int_{z_0}^{z_0+\Delta z} \bar P(z')\,dz'.
\label{eq:dphi_def_main}
\end{equation}

{\it Variance growth and the fluctuation-dissipation connection}---In a normal bulk liquid, $\bar P(z)$ is (approximately) stationary with a finite correlation length $\xi$ along $z$. This means that, in the bulk, the statistical properties of $\bar P(z)$ are nearly invariant under translations along $z$: $\langle \bar P(z)\rangle$ is approximately independent of $z$, and its covariance, $C_p(\zeta)\equiv \langle \bar P(z)\,\bar P(z+\zeta)\rangle$ depends primarily on the separation $\zeta$, with correlations decaying over a finite length $\xi$.

Taking the variance of the potential difference in Eq.~\eqref{eq:dphi_def_main}, assuming statistical homogeneity in the bulk, and rewriting the resulting double integral in terms of the separation $\zeta$ yields
\begin{equation}
\mathrm{Var}\!\big[\Delta\bar\phi(\Delta z)\big]
=\frac{1}{\varepsilon_0^2}\int_{-\Delta z}^{\Delta z}\!(\Delta z - |\zeta|)\,C_p(\zeta)\,d\zeta.
\label{eq:var_general_cont_main}
\end{equation}
The triangle kernel $(\Delta z - |\zeta|)$ counts how many pairs of points separated by $\zeta$ fit within the integration window of length $\Delta z$: it peaks at $\zeta=0$ and vanishes at $|\zeta|=\Delta z$. Because $C_p(\zeta)$ is appreciable only over a finite correlation length $\xi$, the integral is dominated by $|\zeta|\lesssim \xi$. When $\Delta z \gg \xi$, nearly all correlated pairs lie well inside the window, so $(\Delta z-|\zeta|)\approx \Delta z$ throughout the dominant region and $\mathrm{Var}[\Delta\bar\phi(\Delta z)]\propto \Delta z$. The variance of the plane-averaged potential difference thus grows linearly with separation.

This unbounded growth reflects the fact that the uniform plane mode is an unscreened fluctuation. The 2D periodicity forces every lateral replica to carry the identical charge fluctuation, leaving no independent surrounding dielectric medium to provide lateral screening. In a nonperiodic system, by contrast, neighboring regions would screen such a fluctuation and the variance remains bounded (see Control Geometries below).

For an exponential covariance $C_p(\zeta)=C_p(0)\,e^{-|\zeta|/\xi}$, which captures the essential feature of a single correlation length, the integral evaluates to (see Supplemental Material)
\begin{equation}
\mathrm{Var}\!\big[\Delta\bar\phi(\Delta z)\big]
= S\!\left[\Delta z - \xi\!\left(1-e^{-\Delta z/\xi}\right)\right],
\label{eq:Vexact}
\end{equation}
where $S = \varepsilon_0^{-2}\!\int\! C_p(\zeta)\,d\zeta$ is the asymptotic variance slope, reflecting the Wiener-process character of the cumulative potential.

The large-$\Delta z$ behavior, thus, depends only on the integrated polarization covariance, which for the exponential model evaluates to $2\xi\,C_p(0)$. At small distances $\Delta z\ll\xi$, the variance grows quadratically $\sim S\,\Delta z^2/2\xi$, then crosses over near $\Delta z\sim 2\xi$, and approaches linear growth $S\,\Delta z$ for $\Delta z\gg\xi$, i.e.\ a Wiener process in $z$.

The slope $S$ can be expressed entirely in terms of the static dielectric constant $\varepsilon$, without reference to the microscopic parameters $C_p(0)$ and $\xi$.

The total free energy of the plane-averaged polarization profile $\bar P(z)$ has two contributions: a short-range (material) cost $F_{\mathrm{mat}}=\frac{A}{2\chi}\int dz\,\bar P(z)^2$, where $\chi=\varepsilon_0(\varepsilon-1)$ is the static susceptibility, and an electrostatic field energy $F_{\mathrm{el}}=\frac{A}{2\varepsilon_0}\int dz\,\bar P(z)^2$, the latter following from $\bar E=-\bar P/\varepsilon_0$ [Eq.~\eqref{eq:Efield_main}] (see Supplemental Material). Their sum is
\begin{equation}
    F[\bar P]=\frac{A}{2\chi_L}\int dz\,\bar P(z)^2,
\end{equation}
where $\chi_L^{-1}=\chi^{-1}+\varepsilon_0^{-1}$ defines the longitudinal susceptibility $\chi_L=\varepsilon_0(\varepsilon-1)/\varepsilon$~\cite{Jackson1998,LandauLifshitzEDCM}. The susceptibility is \emph{longitudinal} because charge-producing polarization fluctuations incur both a material and an electrostatic energy cost, the latter acting as a depolarization penalty that reduces the effective susceptibility from $\chi$ to $\chi_L<\chi$.

This quadratic free energy implies a local Gaussian response $\chi(z{-}z')=(\chi_L/A)\,\delta(z{-}z')$, so the fluctuation-dissipation theorem gives $C_p(\zeta)=k_BT\,\chi(\zeta)$ and hence
\begin{equation}
    \int\! d\zeta\,C_p(\zeta)=\frac{k_BT\,\chi_L}{A}.
\end{equation}
Equating this with the definition $S = \varepsilon_0^{-2} \int\! C_p(\zeta)\,d\zeta$ from Eq.~\eqref{eq:Vexact} yields
\begin{equation}
S =\frac{k_BT}{\varepsilon_0A}\,\frac{\varepsilon-1}{\varepsilon}
\equiv \frac{s_0}{A},
\label{eq:S_main}
\end{equation}
where $s_0 \equiv k_BT(\varepsilon{-}1)/(\varepsilon_0\varepsilon)$ is the area-independent prefactor of the variance slope.

For high-$\varepsilon$ liquids $s_0\simeq k_BT/\varepsilon_0 \approx 0.47~\mathrm{V}^2\mathrm{nm}$ at 300\,K; it depends only on $\varepsilon$ and $T$, not on the force field or on $\xi$ and $C_p(0)$ individually. The microscopic details enter only through the short-range crossover in Eq.~\eqref{eq:Vexact}, which is bounded and does not affect the divergent growth.

In a cell of length $L$ with $\bar\phi(0) = \bar\phi(L)$ (periodic boundaries along $z$ or two electrodes held at the same potential), the linear growth of the variance in Eq.~\eqref{eq:Vexact} is replaced by a Brownian-bridge parabola:
\begin{equation}
\mathrm{Var}\!\big[\Delta\bar\phi(\Delta z)\big]
= S\!\left[\Delta z\!\left(1-\frac{\Delta z}{L}\right) - \xi\!\left(1-e^{-\Delta z/\xi}\right)\right].
\label{eq:bridge}
\end{equation}
Equations~\eqref{eq:Vexact}, \eqref{eq:S_main} and \eqref{eq:bridge} give a parameter-free prediction for the variance at any $\Delta z$, with the $1/A$ scaling entering through $S$.

{\it Control geometries}---To confirm that the divergence is specific to 2D-periodic replication and not a bulk property, we consider two nonperiodic controls: a purely 1D stack of dipolar sheets (no lateral replication) and a fully 3D nonperiodic medium. In both cases the potential variance converges to a finite plateau with distance, because the relevant electrostatic kernels are summable without the system-spanning $\mathbf{q}{=}0$ mode that 2D periodicity injects. Details are given in the Supplemental Material.

{\it Molecular dynamics test and variance-bound}---We now test the analytic predictions of Eqs.~\eqref{eq:Vexact}, \eqref{eq:S_main}, and \eqref{eq:bridge} against explicit molecular-dynamics simulations of water and estimate the lateral cell areas needed to keep the artifact below a specified tolerance.

We perform classical MD with LAMMPS~\cite{Plimpton1995}, using the flexible TIP3P water model with the Ewald‑optimized nonbonded parameters of Price and Brooks (Model B)~\cite{PriceBrooks2004JCP} and the standard TIP3P intramolecular geometry and flexibility of Jorgensen \emph{et al.}~\cite{Jorgensen1983TIP3P}. Long-range electrostatics were treated with the particle–particle particle–mesh (PPPM) Ewald solver, and simulations were run in the NVT ensemble at 300~K. We simulated cells of cross section $A=1\times1$ and $2\times2$~nm$^2$ and length $L\approx50$~nm, collecting 1~ns of production data after 0.5~ns of equilibration. Statistical convergence was verified by block averaging over four equal sub-intervals, which showed consistent slope and offset parameters within the reported uncertainties. The effective dielectric constant of flexible TIP3P under these conditions is $\varepsilon\approx 82$~\cite{PriceBrooks2004JCP}.

For each snapshot we form the plane-averaged potential $\bar\phi(z)$ from the $\mathbf{q}=0$ Green's-function sum [Eq.~\eqref{eq:phi_general_rho}] and compute
$\mathrm{Var}[\Delta\bar\phi(\Delta z)]
= \langle\,[\Delta\bar\phi(z_0,\Delta z)]^2\,\rangle_{z_0,\text{time}}$.
Figure~\ref{fig:main}(c) compares the MD variance directly with the parameter-free analytic model of Eqs.~\eqref{eq:Vexact}, \eqref{eq:S_main}, and \eqref{eq:bridge}, using only $\varepsilon=82$ and $\xi=0.3$~nm. The agreement is excellent across both cell cross sections and both geometries, with no fitted parameters.

{\it Discussion}---The growth of plane‑averaged potential fluctuations in 2D‑periodic slabs is entirely carried by the uniform plane mode ($\mathbf{q}{=}0$), the fluctuation analogue of the Ewald surface/dipole ($k{=}0$) term~\cite{Ballenegger2014}. Slab-corrected 3D Ewald and true 2D Ewald schemes remove interactions between periodically repeated slabs in $z$ but do not eliminate this mode, which arises from \emph{lateral} periodicity. At any finite lateral area it generates either a linearly increasing or a Brownian-bridge contribution with amplitude $S \propto 1/A$ [Eq.~\eqref{eq:S_main}], whereas nonuniform modes vanish under full-plane averaging or contribute only a bounded background.

A variance that grows without bound with distance cannot represent a physical bulk property. The divergence arises because the uniform plane mode is unscreened under 2D periodicity, making it a boundary-condition artifact that propagates into free energies, biasing solvation, charging, and interfacial estimates even when mean profiles appear converged.

The parameter-free agreement with MD [Fig.~\ref{fig:main}(c)] suggests a simple diagnostic for existing simulations: compare $\mathrm{Var}[\Delta\bar\phi(\Delta z)]$ computed from $\bar\phi(z)$ with the analytic prediction [Eq.~\eqref{eq:bridge}] to assess how much of the observed variance is carried by the $\mathbf{q}{=}0$ artifact.

More generally, when the potential is averaged over a lateral measurement window of area $A_m$ smaller than the 2D-periodic cell area $A$, nonuniform modes $\mathbf{q}{\neq}0$ contribute in addition to the uniform plane mode. Because each nonuniform mode is screened—its Green's function decays exponentially in $z$ [Eq.~\eqref{eq:phi_Gneq0}]—their combined contribution is a bounded, $\Delta z$-independent offset $c_0/A_m$, where $c_0 > 0$ is an intensive constant. Unlike the unscreened uniform $\mathbf{q}{=}0$ artifact, the nonuniform $\mathbf{q}{\neq}0$ contribution is physical: it reflects genuine local potential roughness from finite-wavelength charge fluctuations and is present in nonperiodic systems as well.

The two contributions are additive, and their distinct scaling—$c_0/A_m$ for the physical part versus $S \propto 1/A$ for the artifact—provides a route to separate them: varying $A$ at fixed $A_m$ changes only the $\mathbf{q}{=}0$ terms, while varying $A_m$ at fixed $A$ isolates $c_0$. For full-cell averages ($A_m{=}A$), the $\mathbf{q}{\neq}0$ contribution vanishes by construction.

A direct estimate of the required lateral cell area follows from demanding that the RMS artifact stay below a threshold $\delta\phi$. The Brownian-bridge variance at distance $\Delta z$ from the electrode is $S\,\Delta z(1-\Delta z/L)$, so the RMS condition $\sqrt{S\,\Delta z(1-\Delta z/L)}<\delta\phi$ gives a minimum area
\begin{equation}
    A_{\rm min}\ =\ \frac{s_0\,\Delta z}{\delta\phi^2}\!\left(1-\frac{\Delta z}{L}\right),
\label{eq:Amin}
\end{equation}
where $s_0\simeq k_BT/\varepsilon_0 \approx 0.47~\mathrm{V}^2\mathrm{nm}$ at 300\,K for high-$\varepsilon$ liquids [Eq.~\eqref{eq:S_main}]; for a single interface ($L\to\infty$) the bridge factor is unity. Taking $\Delta z{=}1~\mathrm{nm}$ (a typical reactive-layer thickness), we find: for $\delta\phi{=}0.1$~V (comparable to electrochemical activation barriers), $A_{\rm min}\approx 47~\mathrm{nm}^2$ ($\ell\approx 7$~nm), feasible in classical MD but demanding for \emph{ab initio} simulations; for $\delta\phi{=}0.5$~V (order of an electrode potential window), $A_{\rm min}\approx 1.9~\mathrm{nm}^2$ ($\ell\approx 1.4$~nm), within reach of \emph{ab initio} MD. These estimates depend only on $s_0$ and are therefore independent of the force field.

{\it Conclusion}---
Two-dimensional periodicity injects an unscreened uniform plane mode that converts the plane-averaged polarization into a cumulative potential—the essence of the \emph{$q_{\parallel}$-mode catastrophe}. In 2D-periodic slabs this mode produces potential fluctuations whose variance grows linearly or parabolically with depth, with amplitudes that decay only as $A^{-1/2}$ for the RMS. Without 2D-periodic replication, whether the lateral extent is finite or infinite, fluctuations remain bounded. The divergence is therefore a boundary-condition artifact, not emergent interfacial physics, and it can inflate kinetic and electrostatic-response estimates, especially in small-area polar slabs. Comparing the analytic model of Eqs.~\eqref{eq:Vexact}, \eqref{eq:S_main}, and \eqref{eq:bridge} with simulation data offers a practical diagnostic for identifying and quantifying this artifact.

\acknowledgments
The work was supported by the U.S. Department of Energy, Office of Science, Basic Energy Sciences, under Award DE-SC0019330, and by IPAM at UCLA under NSF Grant DMS-1925919.

\bibliographystyle{apsrev4-2}
\bibliography{references}

\section*{Supplementary Material}

\subsection*{General plane-averaged potential including free charge}
Under two-dimensional periodic boundary conditions, the total plane-averaged charge density decomposes as $\bar \rho(z)=\bar \rho_f(z)-\partial_z \bar P(z)$, where $\bar \rho_f$ is the plane-averaged free charge and $\bar P(z)$ the $z$-component of the plane-averaged polarization. Substituting into the Green's-function solution [Eq.~\eqref{eq:phi_general_rho} of the main text] and integrating by parts yields the general relation
\begin{equation}
\begin{aligned}
\bar\phi(z)
&= -\frac{1}{2\varepsilon_0}\int |z-z'|\,\bar \rho_f(z')\,dz' \\
&\quad + \frac{1}{2\varepsilon_0}\int \bar P(z')\,\mathrm{sgn}(z-z')\,dz'
     \;+\;E_0 z + \phi_0,
\end{aligned}
\label{eq:phi_general_P}
\end{equation}
showing that free charge gives piecewise-linear contributions and polarization gives potential jumps. In the main text we focus on the case $\bar\rho_f=0$ relevant to bulk water.

\subsection*{Energy of Polarization}
To derive an equation for the free energy of the plane-averaged polarization $\bar P(z)$, we separate the free energy into a short-range materials free energy $F_{\mathrm{mat}}$ and the electrostatic field energy $F_{\mathrm{el}}$.

To obtain an expression for the short-range material free energy, 
we introduce an external longitudinal field $\bar E_{\mathrm{ext}}(z)$ conjugate to $\bar P(z)$, so that the coupling is $-A\int dz\,\bar E_{\mathrm{ext}}(z)\bar P(z)$. In linear response, $\bar P=\chi \bar E_{\mathrm{ext}}$, i.e.\ $\bar E_{\mathrm{ext}}(\bar P)=\bar P/\chi$. The reversible work density to build the polarization from $0$ to $\bar P$ is
$f_{\mathrm{mat}}(\bar P)=\int_0^{\bar P} dP'\,\bar E_{\mathrm{ext}}(P')=\bar P^2/(2\chi)$, which yields Eq.~\eqref{eq:SI_Fmat} upon integrating over $z$. Hence, the material (short-ranged/entropic) free-energy cost for a longitudinal polarization profile, $\bar P(z)$, is
\begin{equation}
F_{\mathrm{mat}}[\bar P]=\frac{A}{2\chi}\int_0^L dz\, \bar P(z)^2,
\label{eq:SI_Fmat}
\end{equation}
where $\chi$ is the static susceptibility (SI units F/m), related to the static dielectric constant
$\varepsilon$ by $\varepsilon=1+\chi/\varepsilon_0$ (Ref.~\cite{LandauLifshitzEDCM}).
For the plane-averaged ($q_\parallel=0$) case with negligible plane-averaged free charge, Gauss' law implies
the longitudinal displacement $\bar D(z)=\varepsilon_0\bar E(z)+\bar P(z)$ is $z$-independent. For fluctuations
about the mean in a periodic cell with the $k_z=0$ mode removed (or between constant-potential electrodes),
one has $\delta\bar D=0$, giving
\begin{equation}
\bar E(z)=-\frac{\bar P(z)}{\varepsilon_0}.
\label{eq:SI_E_from_p}
\end{equation}

The electrostatic field energy is
\begin{equation}
F_{\mathrm{el}}=\frac{\varepsilon_0 A}{2}\int_0^L dz\, \bar E(z)^2
\label{eq:SI_Fel}
\end{equation}
(Ref.~\cite{Jackson1998}). Using Eq.~\eqref{eq:SI_E_from_p} in Eq.~\eqref{eq:SI_Fel} gives
$F_{\mathrm{el}}[\bar P]=\frac{A}{2\varepsilon_0}\int_0^L dz\,\bar P(z)^2$. Thus the total quadratic free energy is
\begin{equation}
F[\bar P]
=
\frac{A}{2}\left(\frac{1}{\chi}+\frac{1}{\varepsilon_0}\right)\int_0^L dz\, \bar P(z)^2
\equiv
\frac{A}{2\chi_L}\int_0^L dz\, \bar P(z)^2,
\label{eq:SI_Ftot}
\end{equation}
which defines the longitudinal susceptibility
\begin{equation}
\chi_L^{-1}=\chi^{-1}+\varepsilon_0^{-1}
\quad\Longrightarrow\quad
\chi_L=\frac{\chi\,\varepsilon_0}{\chi+\varepsilon_0}
=\varepsilon_0\,\frac{\varepsilon-1}{\varepsilon}.
\label{eq:SI_chiL}
\end{equation}
Equation~\eqref{eq:SI_chiL} expresses \emph{longitudinal screening}: charge-producing polarization
fluctuations are reduced by the electrostatic energy penalty (Ref.~\cite{LandauLifshitzEDCM}).

\subsection*{Derivation of the analytic variance for exponential covariance}

We derive the closed-form result Eq.~\eqref{eq:Vexact} of the main text. For an exponential covariance $C_p(\zeta)=C_p(0)\,e^{-|\zeta|/\xi}$, the triangle-kernel integral [Eq.~\eqref{eq:var_general_cont_main}] becomes
\begin{equation}
V(\Delta z) = \frac{2C_p(0)}{\varepsilon_0^2}\int_0^{\Delta z}(\Delta z - \zeta)\,e^{-\zeta/\xi}\,d\zeta.
\end{equation}
Evaluating the two terms separately,
$\int_0^{\Delta z}\Delta z\,e^{-\zeta/\xi}\,d\zeta = \xi\,\Delta z\,(1-e^{-\Delta z/\xi})$
and
$\int_0^{\Delta z}\zeta\,e^{-\zeta/\xi}\,d\zeta = \xi^2(1-e^{-\Delta z/\xi}) - \xi\,\Delta z\,e^{-\Delta z/\xi}$,
yields
\begin{equation}
\begin{aligned}
V(\Delta z) &= \frac{2\xi\,C_p(0)}{\varepsilon_0^2}\!\left[\Delta z - \xi\!\left(1-e^{-\Delta z/\xi}\right)\right]\\
&= S\!\left[\Delta z - \xi\!\left(1-e^{-\Delta z/\xi}\right)\right],
\end{aligned}
\end{equation}
where $S = 2\xi\,C_p(0)/\varepsilon_0^2$ is the asymptotic variance slope [Eq.~\eqref{eq:S_main}].

\emph{Limiting behavior.} For $\Delta z\ll\xi$, expanding the exponential gives $V(\Delta z)\approx S\,\Delta z^2/(2\xi)$ (quadratic growth). For $\Delta z\gg\xi$, $V(\Delta z)\approx S\,\Delta z - S\,\xi = S\,\Delta z - \sigma_0^2$ (linear growth with negative intercept), where
\begin{equation}
\sigma_0^2 \equiv S\,\xi = \frac{1}{\varepsilon_0^2}\int_{-\infty}^{\infty}\!|\zeta|\,C_p(\zeta)\,d\zeta \;>\; 0.
\label{eq:SI_sigma0_def}
\end{equation}
The crossover from quadratic to linear occurs at $\Delta z\sim 2\xi$, which for water ($\xi\approx 0.3$~nm) gives $\Delta z_{\rm cross}\approx 0.6$~nm, consistent with Fig.~\ref{fig:main}(c). Note that $V(0)=0$ exactly; the negative intercept $-\sigma_0^2$ of the linear asymptote is not a physical prediction but reflects the initial quadratic ramp. For water at 300\,K with $s_0 = A\,S \simeq 0.47~\mathrm{V}^2\mathrm{nm}$, the intensive offset is $\tilde\sigma_0^2 \equiv A\,\sigma_0^2 \approx s_0\,\xi \approx 0.14~\mathrm{V}^2\mathrm{nm}^2$. A Gaussian covariance yields $\sigma_0^2 = S\,\xi\sqrt{2/\pi}\approx 0.80\,S\,\xi$; the relation $\sigma_0^2\approx S\,\xi$ is robust to the shape of $C_p$.

\emph{Generality.} For an arbitrary short-ranged $C_p(\zeta)$, the general asymptotic expansion is $V(\Delta z) = S\,\Delta z - \sigma_0^2 + O(e^{-\Delta z/\xi})$, where $S=\varepsilon_0^{-2}\int C_p(\zeta)\,d\zeta$ and $\sigma_0^2$ are given by the zeroth and first absolute moments of $C_p$ [Eqs.~\eqref{eq:S_main} and \eqref{eq:SI_sigma0_def}]. The exponential model captures this behavior exactly with a single parameter $\xi$.

Since $C_p \propto 1/A$ (the plane-averaged covariance decreases with lateral area), both $S$ and $\sigma_0^2$ scale as $1/A$. Writing $\tilde\sigma_0^2 \equiv A\,\sigma_0^2$ for the intensive (area-rescaled) offset, the $\mathbf{q}{=}0$ contribution to the variance approaches $S\,\Delta z - \tilde\sigma_0^2/A$ for $\Delta z\gg\xi$, so the short-range correction vanishes along with $S$ as $A\to\infty$. When the potential is averaged over a measurement window $A_m < A$, nonuniform modes ($\mathbf{q}{\neq}0$) enter in addition. Each such mode is governed by the exponentially decaying kernel of Eq.~\eqref{eq:phi_Gneq0} and contributes a bounded, $\Delta z$-independent variance that scales as $1/A_m$ (independent of $A$). The distinct scaling with $A$ versus $A_m$ allows the two contributions to be separated: the $\mathbf{q}{=}0$ terms ($S$ and $\tilde\sigma_0^2/A$) are boundary-condition artifacts that vanish as $A\to\infty$, while $c_0/A_m$ is the $\mathbf{q}{\neq}0$ contribution that is independent of the periodic cell size.

\subsection*{Connection to molecular parameters and the Kirkwood factor}

For a fluid of permanent dipoles of magnitude $\mu$ and number density $n$, the Debye susceptibility is
$\chi_D=n\mu^2/(3k_BT)$ (Refs.~\cite{DebyePolarMolecules,LandauLifshitzEDCM}).
Local orientational correlations are commonly summarized by the Kirkwood correlation factor $g_K$,
leading to $\chi \approx g_K\chi_D$ (Refs.~\cite{Kirkwood1939,FrohlichDielectrics,ZhangSprik2016}).
Using $\chi \approx g_K\chi_D$ with $\chi_D=n\mu^2/(3k_BT)$ and $\chi_L=\varepsilon_0(\varepsilon-1)/\varepsilon$ in
$S=\frac{k_BT\chi_L}{A\varepsilon_0^2}$ [Eq.~\eqref{eq:S_main}], and inserting $S$ into the Brownian-bridge form
$\mathrm{Var}[\Delta\bar\phi(\Delta z)]=S\,\Delta z(1-\Delta z/L)$ [Eq.~\eqref{eq:bridge}], yields the equivalent molecular form
\begin{equation}
\mathrm{Var}\!\left[\Delta\bar{\phi}(\Delta z)\right]
\approx
\frac{1}{3\varepsilon_0^2A}\,\frac{g_K\,n\mu^2}{\varepsilon}\;
\Delta z\left(1-\frac{\Delta z}{L}\right).
\label{eq:SI_gk_form}
\end{equation}
Equation~\eqref{eq:SI_gk_form} highlights the combination $g_K/\varepsilon$: correlations enhance
polarization fluctuations (via $g_K$) while longitudinal electrostatics suppress charge-producing fluctuations
(via $\varepsilon$). In practice, computing $\varepsilon$ directly from dipole fluctuations in periodic simulations
provides a consistent route that avoids double counting correlations (Refs.~\cite{deLeeuwI1980,Neumann1983}).

\subsection*{Control geometries}

The Wiener/Brownian-bridge growth in Eqs.~\eqref{eq:Vexact}--\eqref{eq:bridge} arises from the uniform plane mode injected by lateral replication (2D periodicity). To emphasize that this behavior is not an intrinsic bulk property, we contrast it with two nonreplicated controls in which the potential variance approaches a finite plateau with distance.

\textit{1D stack of dipolar sheets}---Consider a semi-infinite 1D stack of coarse-grained sheets with net dipole moments $\{\boldsymbol{\mu}_n\}_{n\ge 1}$ located on the symmetry axis at positions $z_n=na$ (no lateral replication). We evaluate the on-axis potential at the location of sheet $m$, i.e.\ at $z_m=ma$. The electrostatic potential of a point dipole is $\phi(\mathbf r)=(4\pi\varepsilon_0)^{-1}\,\boldsymbol{\mu}_n\cdot(\mathbf r-\mathbf r_n)/|\mathbf r-\mathbf r_n|^3$. On the symmetry axis, $\mathbf r-\mathbf r_n=(z-z_n)\hat z$, so \emph{only the longitudinal component} $\mu_{n,z}\equiv\boldsymbol{\mu}_n\cdot\hat z$ contributes; transverse components drop out identically. Defining $p_n\equiv \mu_{n,z}$, the on-axis potential at site $m$ is
\begin{equation}
\phi_m
=\alpha\sum_{n\ge 1,\,n\neq m}\frac{\mathrm{sgn}(m-n)}{|m-n|^{2}}\,p_n,
\qquad
\alpha=\frac{1}{4\pi\varepsilon_0 a^{2}}.
\label{eq:phi_1d}
\end{equation}
The key point is that the \emph{squared} kernel decays as $1/|m-n|^4$, which is summable; hence the variance cannot grow without bound with $m$.

Assume $\{p_n\}$ are independent and identically distributed with $\langle p_n\rangle=0$ and $\mathrm{Var}(p_n)=\sigma_p^2=\langle p_n^2\rangle$. Then cross terms vanish and
\begin{align}
\mathrm{Var}[\phi_m]
&=\alpha^2\sum_{n\neq m}\frac{\mathrm{Var}(p_n)}{|m-n|^4}
=\alpha^2\sigma_p^2\sum_{n\neq m}\frac{1}{|m-n|^4}.
\end{align}
Splitting the sum into $n<m$ and $n>m$, and setting $r=|m-n|$, gives
\begin{equation}
\begin{aligned}
\mathrm{Var}[\phi_m]
&=\alpha^2\sigma_p^2\!\left[\sum_{r=1}^{m-1}r^{-4}+\sum_{r=1}^{\infty}r^{-4}\right]\\
&=\alpha^2\sigma_p^2\!\left[H^{(4)}_{m-1}+\zeta(4)\right],
\end{aligned}
\label{eq:var_1d_exact}
\end{equation}
where $H^{(4)}_{m-1}=\sum_{r=1}^{m-1}r^{-4}$ is a generalized harmonic number and $\zeta(4)=\sum_{r=1}^{\infty}r^{-4}=\pi^4/90$ is the Riemann zeta function. Using the large-$m$ expansion $H^{(4)}_{m-1}=\zeta(4)-\frac{1}{3m^{3}}+O(m^{-4})$, one finds
\begin{equation}
\mathrm{Var}[\phi_m]
=\alpha^2\sigma_p^2\left[2\zeta(4)-\frac{1}{3m^{3}}+O(m^{-4})\right].
\label{eq:var_1d_asymp}
\end{equation}
Thus the variance saturates to the finite plateau $\mathrm{Var}[\phi_m]\to 2\alpha^2\sigma_p^2\zeta(4)=\alpha^2\sigma_p^2\pi^4/45$ as $m\to\infty$, approached with a $1/m^3$ tail. If the dipoles are not independent but have short-ranged correlations (e.g.\ $\sum_{r\ge 0}|\mathrm{Cov}(p_n,p_{n+r})|<\infty$), the variance remains bounded and only the overall prefactor is renormalized.

\textit{Fully 3D, nonperiodic medium}---In a fully 3D, nonperiodic geometry (no lateral replication), there is no isolated system-spanning $q_\parallel=0$ mode. Potential fluctuations are generated by genuinely three-dimensional dipolar fields. The far-field potential of a point dipole $\boldsymbol\mu$ at separation $\mathbf r$ is (Ref.~\cite{Jackson1998})
\begin{equation}
\phi(\mathbf r)=\frac{1}{4\pi\varepsilon_0}\,\frac{\boldsymbol\mu\cdot \mathbf r}{r^{3}},
\qquad |\phi(\mathbf r)|\sim r^{-2}.
\end{equation}
For isotropically oriented dipoles, $\langle(\boldsymbol\mu\cdot\mathbf r)^2\rangle=\langle\mu^2\rangle r^2/3$, hence
\begin{equation}
\langle \phi(\mathbf r)^2\rangle
=\frac{\langle\mu^2\rangle}{48\pi^2\varepsilon_0^2}\,\frac{1}{r^{4}}.
\end{equation}
For independent dipoles of number density $n$, the \emph{far-field} contribution to the variance scales as
\begin{equation}
\mathrm{Var}[\phi]_{\mathrm{far}}
\sim n\!\int d^3r\,\langle \phi(\mathbf r)^2\rangle
\propto \int_{r_0}^{\infty}\! \frac{dr}{r^{2}}<\infty,
\end{equation}
where $r_0$ is a microscopic cutoff set by molecular size (or any finite coarse-graining/measurement window). Thus $\mathrm{Var}[\phi]$ (and therefore the variance of any finite-area plane average of the potential) approaches a finite bulk plateau with distance from a boundary, in contrast to the 2D-periodic slab where a discrete uniform plane mode produces Wiener/Brownian-bridge growth.
\end{document}